\documentclass[onecolumn]{nature}
\usepackage{graphicx,color,dcolumn,bm,float,longtable,mathptmx}
\usepackage{amsfonts,amsmath,amssymb}
\usepackage{amsfonts,amsmath,amssymb}

\title{La displacement driven Double-exchange like mediation in Titanium $d_{xy}$ ferromagnetism at the LaAlO$_3$/SrTiO$_3$}
\author{Dorj Odkhuu$^{1,2}$, S. H. Rhim$^{3}$, Dongbin Shin$^1$, and  Noejung Park$^1$}
\begin{document}
\maketitle
\begin{affiliations}
\item Department of Physics, Ulsan National Institute of Science and Technology (UNIST), Ulsan, Republic of Korea\\
\item Department of Physics, Incheon National University, Incheon, Republic of Korea\\
\item Department of Physics and Energy Harvest Storage Research Center, University of Ulsan, Ulsan, Republic of Korea\\
\end{affiliations}
\date{\today}

\begin{abstract}
  The epitaxial atomistic interfaces of two insulating oxides,
  LaAlO$_3$ (LAO)/SrTiO$_3$ (STO), have attracted great interest owing
  to rich emergent phenomena \cite{hwang12:nmat,coey13:mrs,ohtomo04:}
  such as interface metallicity \cite{ohtomo04:,nakagawa06:},
  thickness dependent insulator-metal transition \cite{huijben06:nmat},
  superconductivity \cite{N.Reyren08312007}, ferromagnetism \cite{brinkman.07:natmat},
  and even their coexistence \cite{bert11:nphys,li11:nphys,dikin11:prl}.
  However, the physics origin of ferromagnetic ordering in the $n$-type LAO/STO interface is in debate.
  Here we propose that the polar distortion of La atom can ignite the ferromagnetism
  at the interface even without oxygen vacancy.
  The induced hybridization between La $d_{z^2}$ and O $p_{x,y}$ states
  can mediate double-exchange like interaction between Ti $d_{xy}$ electrons.
  We further suggest that the structural and electrical modification of the outermost surface of LAO or
  switching the polarization direction of ferroelectric overlayers on LAO/STO
  can promote such La displacement.
\end{abstract}

Since the first report of the unexpected conductivity in the interface of two hetero oxide insulators,
LaAlO$_3$/SrTiO$_3$ (LAO/STO),
quite numerous studies have been concentrated on the microscopic origin of the observed metallicity.
While the electronic reconstruction to avoid polar catastrophe
provides the intrinsic source of the two-dimensional electrons \cite{pentcheva09:prl},
the extrinsic mechanisms,
such as oxygen vacancies \cite{siemons07:prl,herranz07:prl},
the defects in LAO surface \cite{slooten:2013,li:2011,bark12}
and cationic disorder\cite{willmott07:prl,kalabukhov09:prl} prevails the actually observed conductivity.
Recently, a possible adjustment of the conductivity through the treatment of the outer surface of the LAO
is now gathering a renewed interest.\cite{slooten:2013,li:2011,bark12,mix14:apl,tra13:advmat}
It was shown that adsorptions of charged species or polarization switching of ferroelectric overlayers
can adjust the overall polarization direction of the whole LAO layer,
which can tune on or off the conductivity at the LAO/STO interface. 

Meanwhile magnetism in the LAO/STO has become an issue of great significance\cite{brinkman.07:natmat}. 
An earlier theoretical work proposed magnetism with charge ordering,
an checkerboard type antiferromagnetism (AFM) \cite{pentcheva06_prb:}.
While observation of such checkerboard-type AFM is still missing,
ferromagnetism in the $n$-type LAO/STO interface has been observed --
evidenced by hysteresis in magnetoresistance\cite{brinkman.07:natmat}.
With this intriguing interface ferromagnetism,
there are still on-going debates on the physics origin.
It has been widely believed that the formation of oxygen vacancies in the interface is indispensable
for magnetic instability\cite{pavlenko12:prb,pavlenko12:prbR,lechermann14:prb}.
On the other hand, other alternative mechanisms have been proposed theoretically,
whose underlying physics is radically different from vacancy related ones:
quasi-one-dimensional band can drive magnetic instability without any coupling
between local moments and conduction electrons\cite{chen13:prl};
ferromagnetic ordering of localized charge via exchange interaction
with the conduction electrons\cite{michaeli12:prl}.
Recently, X-ray magnetic circular dichroism (XMCD) revealed that the interface ferromagnetism is
from the exchange splitting of Ti $d_{xy}$ state
that is hybridized with the O 2$p$ orbitals\cite{lee13:nmat}.

In this article, we propose that ferromagnetism at the $n$-type LAO/STO interface
is possible without an oxygen vacancy at the interface.
A level mixture of O $p_{x,y}$ and La $d_{z^2}$ enables Ti $d_{xy}$ ferromagnetism
by polar distortion of La atoms,
without which other defects are required for ferromagnetism.
To realize the La displacement, various kinds of defects such as an oxygen
vacancy and adsorbates at the outer surface AlO$_2$ of the LAO
layer are considered. Furthermore, we demonstrate tunability
of La distortion by depositing ferroelectric overlayers on LAO/STO
-- a switching on/off of Ti $d_{xy}$ ferromagnetism.

Fig.~~1 systematically explains our mechanism:
Ti $d_{xy}$ ferromagnetism is mediated by hybridized state of La $d_{z^2}$ and O $p_{x,y}$,
which offers additional path for double-exchange like mediation between localized Ti moments.
Relevant orbitals in this mechanism are depicted in Fig.~~1a
for Ti $d_{xy}$, O $2p_{x,y}$, and La $d_{z^2}$,
where the hybridization between La $d_{z^2}$ and O $2p$ is realized
when La moves closer toward the interface.
In perovskite structure, the crystal field under $O_h$ point group
splits five-fold degenerate Ti $d$ states
into two-fold and three-fold degenerate $e_g$ and $t_{2g}$ states.
In tetragonal symmetry, when LAO/STO is formed, $e_g$ further splits into two singlets ($b_1$ and $a_1$),
and $t_{2g}$ states into one doublet ($e=xz,yz$) and one singlet ($b_2=xy$) (Fig.~~1b).
It is this $b_2$ state, or $d_{xy}$ orbital of Ti that
involves occupation of low-lying state in the $n$-type LAO/STO.
To understand ferromagnetism in LAO/STO,
energy levels of Ti $d_{xy}$ and O $2p$ are presented in Fig.~~1c,
which corresponds to the case without any defect. Hence, La $d_{z^2}$ is the high lying state.
If Ti is ferromagnetically aligned with $d_{xy}$ occupation,
hopping from one Ti site to other via hybridization with O site is prohibited due to Pauli principle.
Moreover, the mediation through La site is not likely due to small energy gain
$t^2/\Delta$, where $t$ is the hopping,
$\Delta$ is the energy of La $d_{z^2}$ relative to Ti $d_{xy}$.
In this limit, the ferromagnetically aligned state can be only insulating without any defect.
Thus, metallic ferromagnetism is only possible when oxygen vacancy is introduced
either by having empty state at O or some mixed valency of Ti site.
However, when La moves towards to the interface,
La $d_{z^2}$ will be nearby O $2p$ state,
so these two states form a linear combination, $|{\rm O},p_{xy}\rangle \otimes |{\rm La},d_{z^2}\rangle$ [Fig.~~1d].
Consequently, ferromagnetically aligned Ti  $d_{xy}$ state can thus be stabilized by allowing
hopping between Ti sites via O and La sites, hence ferromagnetic metallicity can be realized.

The atomic configurations of 5 unit cell (u.c.) layers of LAO on 4 u.c. STO with either AlO$_2$- or LaO-termination
are shown in Fig.~~2a. Hereafter, numbers of LAO and STO layers are denoted in subscripts,
and the boundary layers LaO and TiO$_2$ at the interface are labeled as LaO(I) and TiO$_2$(I).
Notably, La displacement depends on the termination.
La atoms move up (down) for the AlO$_2$- (LaO) termination.
The relative displacement of La atom, denoted as $d_{{\rm La}}$, with respect to the corresponding O plane
is schematically presented in Fig.~~2b,
where positive (negative) value of $d_{{\rm La}}$ represents downward (upward) shift of La atom
with respect to the O plane.
As shown in Fig.~~2c,
the $d_{{\rm La}}$ at the boundary LaO(I) layer is substantial for both terminations
although its absolute value decreases monotonically as La goes away from the interface.
Similar results were reported for the AlO$_2$-terminated case \cite{pentcheva09:prl}.

The layer-resolved density of states (LDOS) of the TiO$_2$(I) layer
in (LAO)$_5$/(STO)$_4$ and (LAO)$_{4.5}$/(STO)$_4$ are plotted in Fig.~~2d.
For the AlO$_2$ termination, as with previous study\cite{pentcheva09:prl},
the valence and conduction band consist of O $2p$ and Ti $3d$ state, respectively.
The top of the valence bands O 2$p$ in the surface AlO$_2$ layer shifts
towards the Fermi level as the $n$ in (LAO)$_{n}$/(STO)$_4$ increases,
and eventually meets the bottom of conduction bands Ti 3$d$  of the TiO$_2$ interface at $n$=4.
As a result, the charge transfer takes place from the surface AlO$_2$ layer to the TiO$_2$(I),
well manifested by small occupation just below the Fermi level in Fig.~~2d with no magnetism.
On the other hand,
for the LaO termination, distinctly different features appear:
the interface metallicity near the TiO$_2$(I) layer
leads to exchange-splitting in Ti 3$d$ orbital.
This interface metallicity is mainly from Ti $d_{xy}$ state,
while other Ti $d$ states of TiO$_2$(I) layer are empty.
Directions of La displacements are clearly different in two terminations,
and the positive $d_{{\rm La}}$ (La moving towards the interface) breaks the spin degeneracy of Ti $d_{xy}$.

The role of La displacement is demonstrated in further calculations as depicted in Fig.~~3a,
where La atoms are shifted uniformly while other atoms are fixed in the AlO$_2$-terminated (LAO)$_5$/(STO)$_4$.
The calculated charge and magnetic moment of Ti atom in the TiO$_2$(I) layer are plotted
as a function of $d_{{\rm La}}$ in Figs.~~3b and c, respectively.
Charge of Ti ($\Delta\rho_{Ti}$) is defined as the increment of the Bader charge
with respect to that with $d_{{\rm La}}$=$-$0.3~\AA.
As La atoms shift towards the boundary TiO$_2$-layer,
the occupation of Ti $d_{xy}$ orbital gradually increases.
This serves as a control parameter in ferroelectric polarization:
the polarization up (away from the interface, $P_{\uparrow}$) gives rise to charge depletion and
switching the polarization direction (towards the interface, $P_{\downarrow}$) results
in charge accumulation.
Importantly, magnetism appears when $d_{{\rm La}}$ just passes its equilibrium position
  (left vertical line in Fig.~~3c),
which further increases as La atom moves closer to the interface.
For $d_{{\rm La}}$=0.17~\AA~
(the right vertical dashed-line in Fig.~~3c),
which corresponds to the equilibrium position of La at the LaO(I)
in the LaO-terminated LAO/STO (see Fig.~~2c),
the Ti moment reaches about 0.5 $\mu_{B}$.
We also note that magnetism appears even when $d_{{\rm La}}=0$ with the value about 0.2 $\mu_{B}$.
The LaO termination exhibits similar trends:
the moment and charge increase as La moves toward the interface ($d_{{\rm La}}>0$),
whereas they decrease with the upward displacement of La atom, and eventually vanish
when $d_{{\rm La}}$ approaches to 0.3 ~\AA~ (not shown).

To unveil the electronic structure  more detail,
the orbitally projected DOS (PDOS) of Ti atom at the TiO$_2$(I), La atom at the LaO(I),
and O atom at the TiO$_2$(I)
are plotted in Fig.~~3~d--f, respectively.
For better comparison, those for $d_{{\rm La}}$=$-$0.3~\AA~(non-magnetic)
and $d_{{\rm La}}$=0.3~\AA~(ferromagnetic) are also shown,
which are consistent with that of fully relaxed geometry (Fig.~~2d).
When it is non-magnetic ($d_{{\rm La}}$=$-$0.3~\AA),
conduction bands consist of Ti and La $d$ orbitals while valence bands of O $p$ states,
hence Ti$^{4+}$ {\em or} $3d^{0}$.

In the case of ionic displacement of $d_{{\rm La}}=0.3$~\AA,
Jahn-Teller split $d_{xy}$ shift more downward and is occupied,
while other $t_{2g}$ states ($d_{xz,yz}$) remain unoccupied.
As addressed in Fig.~~3d-f, bands of the
Ti $d_{xy}$, La $d_{z^2}$, and O $p_{x,y}$ states overlap
in the vicinity of the Fermi level giving hybridization of these orbitals.
This hybridization, as illustrated in Fig.~~1, enables double-exchange like mediation of Ti $d_{xy}$ ferromagnetism.
We also emphasize here the electronic structure is consistent with previous experiments:
Ti atom at the interface shows a feature of Ti$^{3+}$ ($3d^{1}$) state
with fractional occupation of $t_{2g}$ orbital states \cite{cen08:nmat};
magnetic moment probed by XMCD is mainly by Ti $d_{xy}$
with hybridization between Ti and O $2p$ \cite{lee13:nmat}.
Moreover, magnetic moment and $\Delta \rho_{Ti}$ of Ti(I) increase
as the cation atom comes closer ($d_{{\rm La}}>0$).

To achieve La displacement,
as illustrated in Fig.~~4~a--f,
the following surface adsorbates and defect configurations at the outermost surface of LAO, the surface of AlO$_2$-layer,
are taken into account:
a. clean-AlO$_2$, b. O-vacancies,  c. H-adsorptions,
d. OH-adsorptions, e. Al-adatoms, and  f. La-adatoms.
Numerous experiments have already demonstrated that
the electronic structure of the interface can be switchable between insulating and conducting phases
by adjusting the outermost LAO surface
such as capping overlayers or adsorptions, as mentioned previously \cite{bi10:apl,xie11:ncom,mix14:apl,tra13:advmat,kim13:advmat}.
The La displacements and spin moments at the boundary layers for these six configurations are
summarized in Fig.~~4b and ~4c, respectively.
Except the clean-AlO$_2$ and the OH-adsorption,
other configurations produce substantial change in the charge polarity of the outermost AlO$_2$ layer.
One notes that $d_{{\rm La}}<0$ for the clean-AlO$_2$ and the OH-adsorption with net zero moment.
However, the oxygen vacancy and the H-adsorption introduce
localized holes near the surface layer,
which we denote as (AlO)$^{+}$ and (AlO$_2$H$_2$)$^{+}$, respectively.
These positive charges of the outermost surface push La cations downwards electrostatically,
which eventually results in Ti $d_{xy}$ ferromagnetism with localized metallicity near the TiO(I).
Similarly, $d_{{\rm La}}>0$ for the Al- and La-adatom,
where the resulting magnetic of the Ti(I) is as large as 0.60 $\mu_{B}$.
We recall here that manipulation of  charge state at the outermost surface of LAO
has been already realized to control the interface metallicity\cite {bark12,xie13:avdmat,bi10:apl},
where magnetism is yet to be explored.

As some adsorbates on the outermost surface gives non-vanishing magnetic moments,
we explore more with depositing ferroelectric BaTiO$_3$ (BTO) overlayer on top of the LAO layer.
Three u.c. thick BTO layers are placed on top of (LAO)$_5$/(STO)$_4$, as shown in Figs.~~5 a and b,
where two polarizations are taken into account, P$_{\uparrow}$ and P$_{\downarrow}$, respectively.
$P_{\uparrow}$ ($P_{\downarrow}$) represents the direction of the ferroelectric polarization of BTO
pointing away from (pointing towards) the interface BTO/LAO.
The atomic coordinates of BTO to be kept fixed in their bulk positions to retain its ferroelectricity,
while other layers are fully relaxed.
More specifically, displacements of Ba and Ti atoms 
relative to the plane containing O in BTO are 0.36~\AA~ and 0.29~\AA, respectively.
The ferroelectric polarization is estimated to be 54.2 $\mu$C/cm$^2$
from Berry phase calculations\cite{resta93,King-Smith:prb},
which is in the range of the experimental values for the strained BTO (50--70 $\mu$C/cm$^2$) \cite{choi04:sci}.
La displacements in different polarization are in opposite directions, i.e.,
$d_{\rm La}<0$ for $P_{\uparrow}$ and $d_{\rm La}>0$ for $P_{\downarrow}$.
This is in accordance with the recent experiments that the direction of internal electric field of LAO layers is determined
by the polarization direction of ferroelectric top layer \cite{tra13:advmat}.

PDOS of the boundary layers LaO and TiO$_2$ for these two opposite polarizations are shown
in Figs.~~5c and d, respectively.
Clearly, PDOS for two polarizations are different.
For $P_{\uparrow}$, the La $d_{z^2}$ at the LaO(I) is far above the Fermi level
and there is no metallicity nearby the interface, thus not to mention of magnetism.
In contrast, for $P_{\downarrow}$, when the polarization is pointing downward,
La atoms move downward, and ferromagnetic metallicity is well addressed with moment of 0.25$\mu_{B}$.
The empty majority Ti $d_{xy}$ state at the TiO$_2$(I) shifts down towards the Fermi level.

To summarize, we propose a possible mechanism for ferromagnetism in the $n$-type LAO/STO interface without any interfacial defects.
As proposed in previous studies,
the interface ferromagnetism can be realized by mixed-valency of Ti $d$ state in the presence of oxygen vacancy.
In our model, even without an oxygen vacancy, additional hopping channel --
a bypass through the hybridized state between La $d_{z^2}$ and O $2p$ --
enables double-exchange like mediation of Ti $d_{xy}$ ferromagnetism.
In this alternative mechanism, the polar distortion of La atom is an essential ingredient,
whose direction of displacement depends on the charge polarity of the outermost surface layer of LAO.
\begin{methods}
The density-functional theory (DFT) calculations were carried out
with the Vienna ab initio simulation package (VASP) \cite{kresse93:prb}.
For the electron exchange correlations,
the PBE-type generalized gradient approximations (GGA) \cite{pbe:96}
were used together with the Hubbard-type on-site Coulomb energy
on Ti 3$d$ orbitals (DFT+U) \cite{liechtenstein95:prb}.
The variations of U do not change qualitatively our conclusion,
and results shown here were calculated with U=5 eV and J=1 eV [20].
As a model geometry, we considered five u.c. layers of LAO
on four u.c. STO
with a vacuum region at least 12~\AA~ between the repeated slabs.
The various test calculations for different thicknesses of LAO and STO layers,
and symmetric LAO/STO/LAO heterostructures have been
also performed to ensure robustness of present calculations.
The dipole corrections were taken into account
to eliminate an artificial electric field
across the slab imposed by the periodic boundary condition.
The experimental lattice constant (3.905~\AA) of STO was used
for the two-dimensional lattice of the LAO/STO slab.
All ionic positions were fully relaxed.
An energy cutoff of 600 eV and a 16$\times$16$\times$1 {\em k}-point grid
were used, whose convergences were ensured.
In BTO/LAO/STO, the in-plane lattice constant of STO, 3.905~{\AA}, is taken,
where optimized lattice constant of BTO along the $z$-direction is 4.491~{\AA}.
\end{methods}

\begin{addendum}
\item   
We thank W. E. Pickett for helpful discussions.
This work was supported
by Basic Science Research Program through the NRF funded
by the Ministry of Education (NRF-2013R1A1A2007910).
SHR is supported by Priority Research Centers Program (2009-0093818) and 
the Basic Science Research Progam (2009-0088216) through
the National Research Foundation (NRF) funded by the Korean Ministry of Education.
\item[Authors contributions]
DO performed all calculations.
NP initiated the work.
DS helped computation details.
DO, SHR, and NP wrote the manuscript.
\item[Correspondence]
Reprints and permissions information is available online at www.nature.com/reprints.
Correspondence and requests for materials\\
should be addressed to NP~(email: noejung@unist.ac.kr)
or SHR~(email: sonny@ulsan.ac.kr).
\item[Competing Interests]
  The authors declare no competing financial interests.
\end{addendum}


\begin{figure}
  \centering
\caption{ 
  {\bf Schematic diagram of double-exchange like mechanism of ferromagentism via La displacement.}
{\bf a.} Schematics for the coupling of Ti $d_{xy}$, La $d_{z^2}$, and O $2p_{x,y}$ orbitals.
{\bf b.} Energy levels of $d$ orbital states in the presence of crystal field.
In $O_h$ group, $d$ orbital splits into two- and three-fold degenerate
$e_g$ and $t_{2g}$ states, which are decomposed further under tetragonal symmetry:
$e_g$ into $b_1=(3r^2-z^2)$ and $a_1=(x^2-y^2)$,
and $t_{2g}$ into $e=(xz,yz)$ and $b_2=(xy)$, respectively.
{\bf c.} The electronic energy level near the low-lying Ti $d_{xy}$ state
at the LAO/STO interface without La distortion,
and the schematics for the forbidden electron itinerancy. 
{\bf d.} Adjusted energy level at the interface, on the La displacement,
and schematics for the double-exchange between Ti 3$d_{xy}$ electrons. 
}
  \label{fig:1}
\end{figure}

\begin{figure}
  \centering
 \caption{
   {\bf The effects of surface termination on La displacement and interface electronic structure.}
   {\bf a.} Optimized atomic structures
   for AlO$_2$ (left)- and LaO (right)-terminated (LAO)$_5$/(STO)$_4$,
   where bottom two unit cell layers of STO are not shown for simplicity.
   Small (red) balls at the vertices of octahedron are oxygen atoms,
   black and blue spheres are La and Sr atoms, respectively.
   Ti atoms are hidden by octahedrons. Atomic planes
   with cations are denoted explicitly with their nominal charge.
   The LaO and TiO$_2$ layers at the boundary are labeled by LaO(I) and TiO$_2$(I), respectively,
   where I stands for interface and S for surface.
   {\bf b.} Schematic representation of La displacements from the corresponding O plane in the LAO.
   Upward (downward) movement toward the surface (interface) is indicated by the negative (positive)
   displacement of La ions ($d_{\rm{La}}$).
   The amplitude of $d_{\rm{La}}$ is multiplied by a factor of two in {\bf a} for better visualization.
   {\bf c.} The La displacements, $d_{\rm{La}}$, of each LaO layer of (LAO)$_5$/(STO)$_4$
   for the AlO$_2$ (black symbol)-   and LaO (red symbol)-termination.
   {\bf d.} LDOS of the TiO$_2$(I) layer for the AlO$_2$ (yellow shaded)-
   and the LaO (red solid line)-terminated (LAO)$_{5}$/(STO)$_{4}$.
   The Fermi level is set to zero for all cases.}
 \label{fig:2}
\end{figure}

\begin{figure}
  \centering
 \caption{
   {\bf Role of La displacement on the interface ferromagnetism}.
   {\bf a.} Optimized atomic structure of AlO$_2$-terminated (LAO)$_{5}$/(STO)$_4$.
   {\bf b.} Charge ($\triangle \rho_{{\rm Ti}}$) and {\bf c.} magnetic moment of the Ti atom at the TiO$_2$(I) layer
   of (LAO)$_{5}$/(STO)$_{4}$ versus the uniform displacement of La ion ($d_{\rm{La}}$) shown schematically in {\bf a}.
   The left and right vertical line denote $d_{{\rm La}}$ of La displacement
   corresponding to fully relaxed geometry of (LAO)$_5$/(STO)$_4$ for the AlO$_2$- and LaO-terminated, respectively.
   In {\bf c.} $\Delta \rho_{\rm Ti}$ when $d_{\rm{La}}$= $-$0.3 {\AA} is taken as reference.
   PDOS for {\bf d.} the Ti $d_{xy}$ at the TiO$_2$(I),
   {\bf e.} La $d_{z^2}$ at the LaO(I), and
   {\bf f.} O $p_{x,y}$ orbitals in the TiO$_2$(I)
   with $d_{\rm{La}}$=$-$0.3 {\AA} (shaded (yellow) area)
   and $d_{\rm{La}}$=0.3 {\AA} (solid line).
   Dashes squares near the Fermi level serve to emphasize orbital hybridization.
     The Fermi level is set to zero. }
\label{fig:3}
\end{figure}

\begin{figure}
  \centering
\caption{
  {\bf La displacement and interface ferromagnetism
    by the structural and electrostatic surface modifications}.
  {\bf a.} Six different surface morphologies. 
  Atoms are represented by the same symbols as in Fig. 2,
  and the small blue balls are hydrogen atoms.
  {\bf b.} The La displacement, $d_{\rm{La}}$, at the LaO(I) and {\bf c.} magnetic moment of Ti atom at the TiO$_2$(I)
  for six surface configurations.}
 \label{fig:4}
\end{figure}

\begin{figure} 
  \centering
  \caption{
    {\bf Tunable interface ferromagnetism by polarization of ferroelectric BaTiO$_3$ overlayer on LAO/STO}.
    Schematic view and the optimized atomic structures of (BTO)$_3$/(LAO)$_3$/(STO)$_2$
    for {\bf a.} Upward-polarization ($P_\uparrow$),
    and {\bf b.} downward-polarization ($P_\downarrow$) of the ferroelectric polarizations of BTO.
    Green sphere represent Ba atoms, while other symbols are the same as in Fig.~~2.
    LDOS of the boundary layers: {\bf c.} LaO(I) and {\bf d.} TiO$_2$(I) for
    the P$_\uparrow$ (yellow shade) and P$_\downarrow$ (red solid line).
    The Fermi level is set to zero. }
 \label{fig:5}
\end{figure}

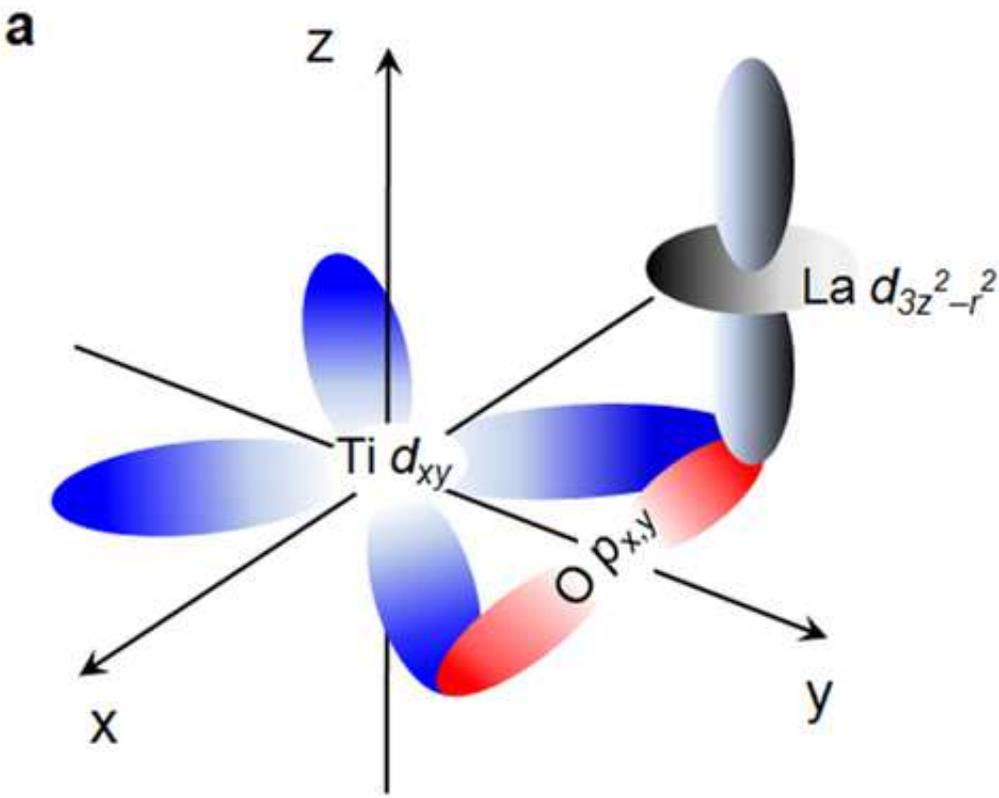
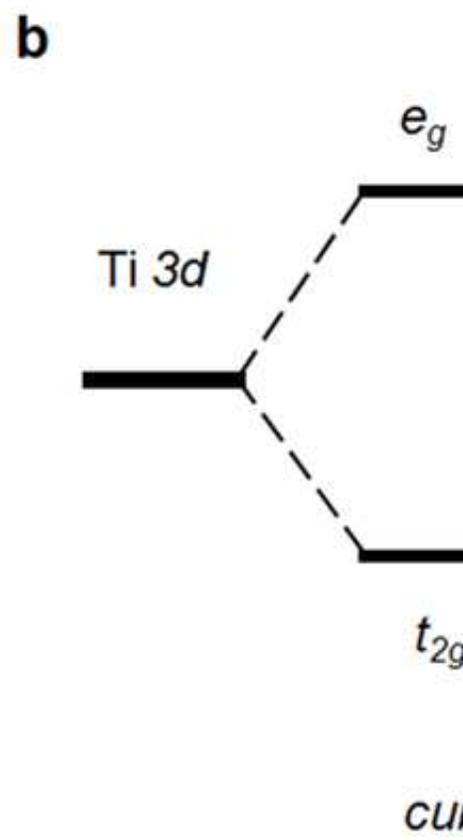
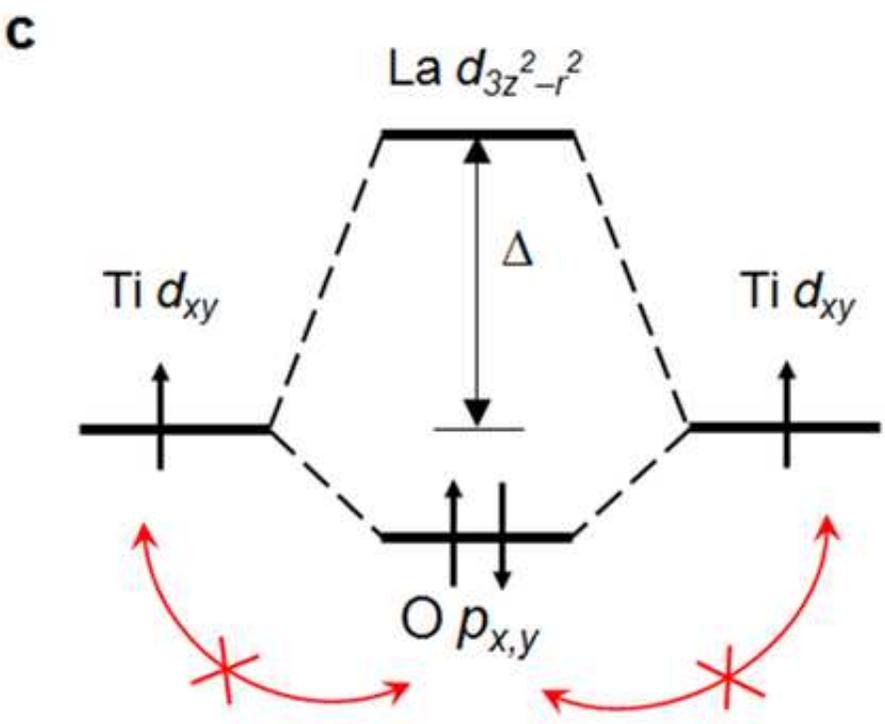
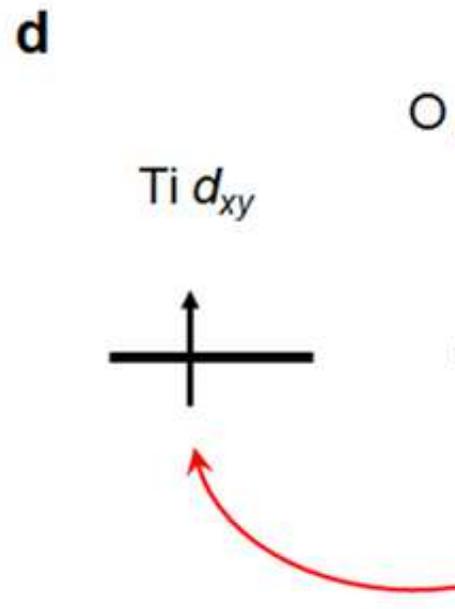
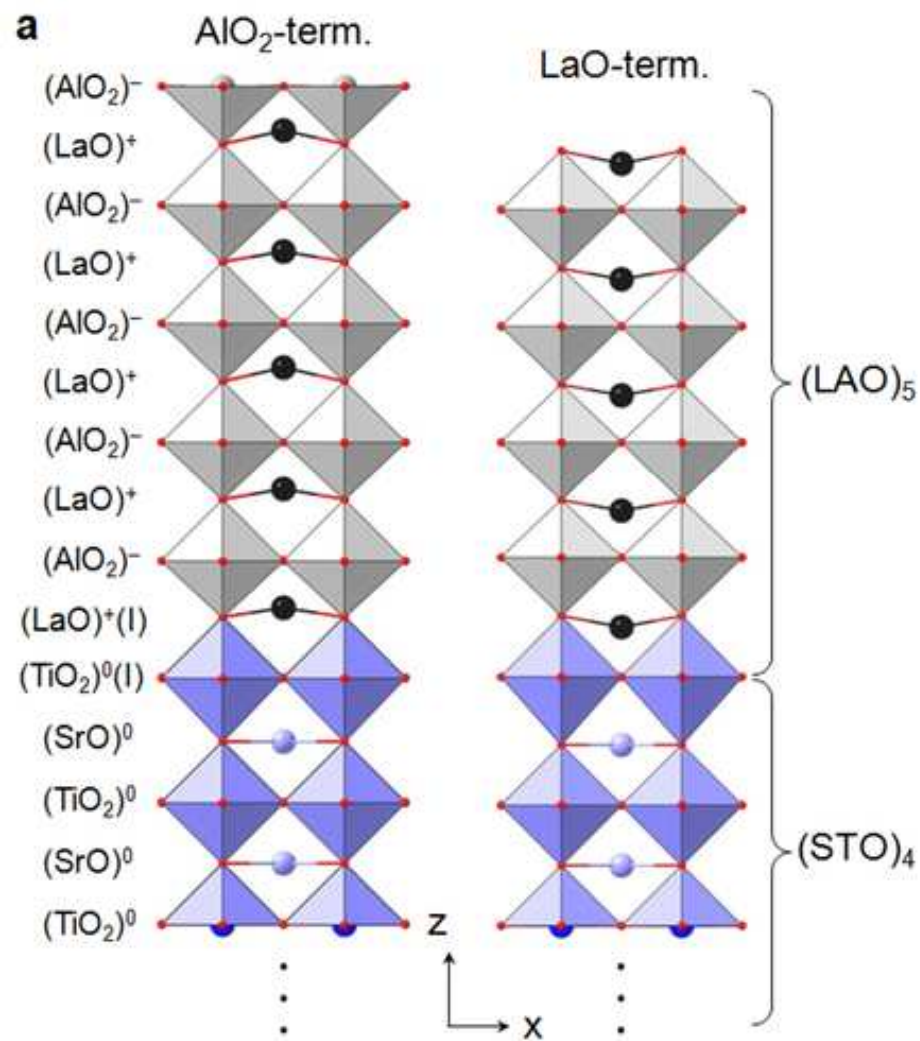
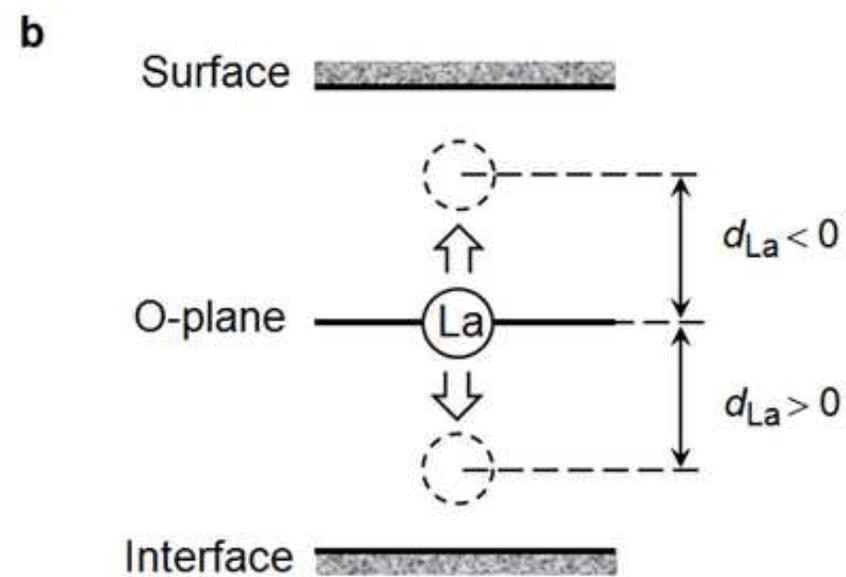
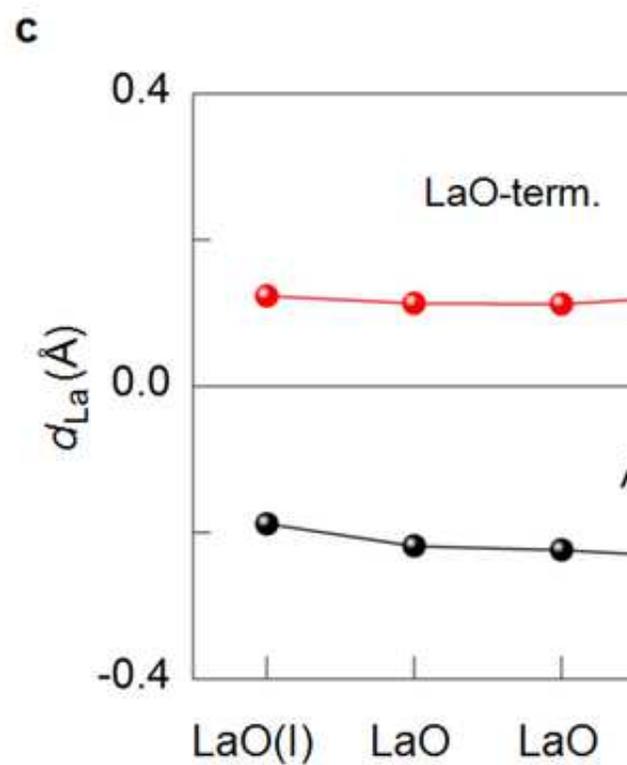
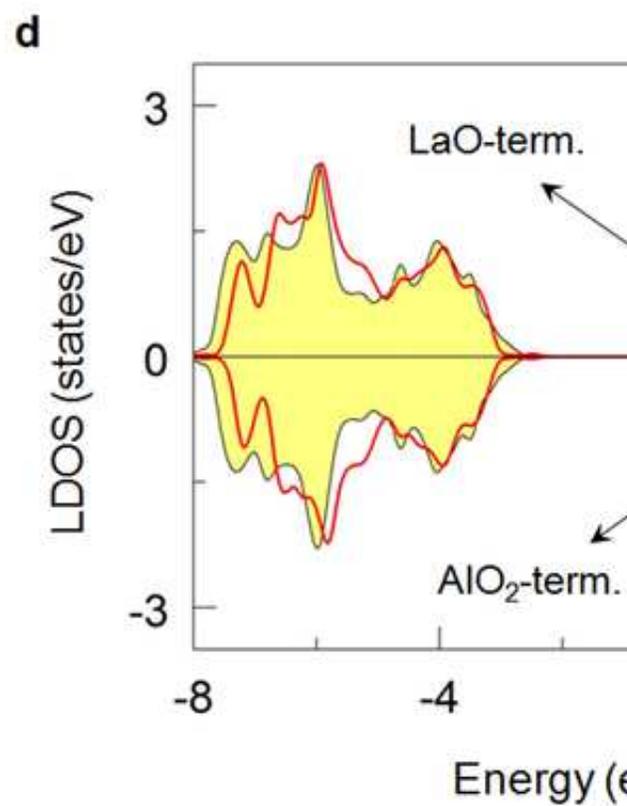
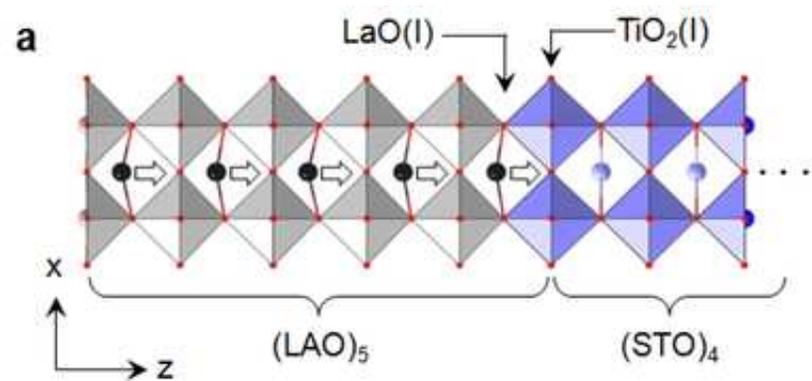
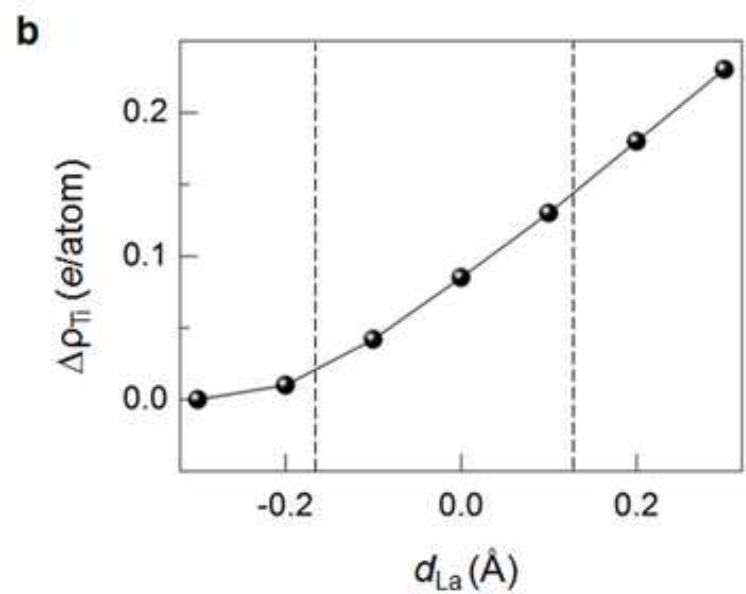
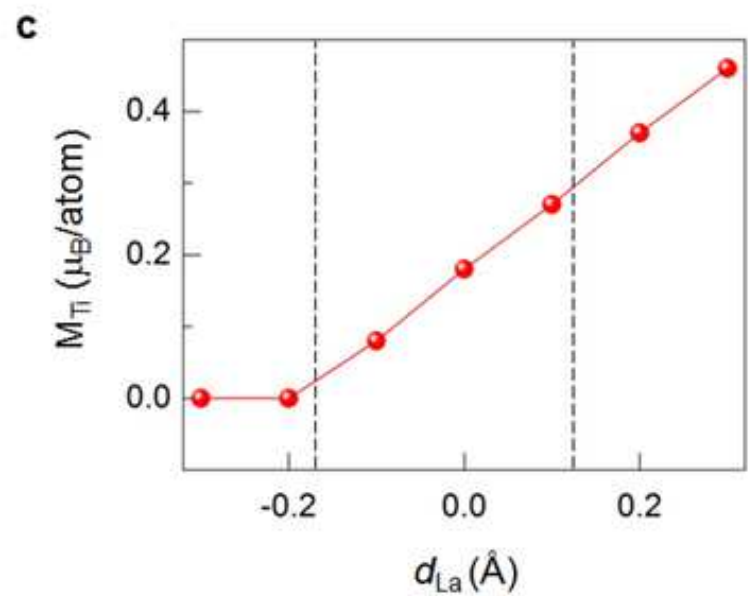
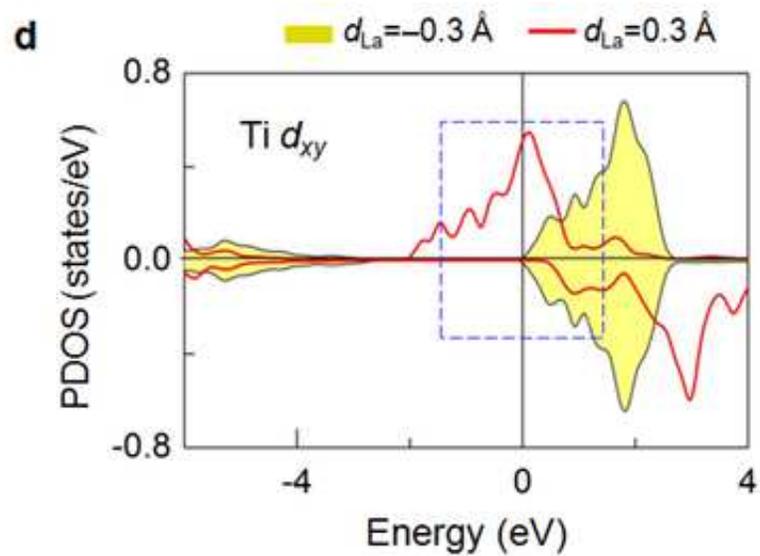
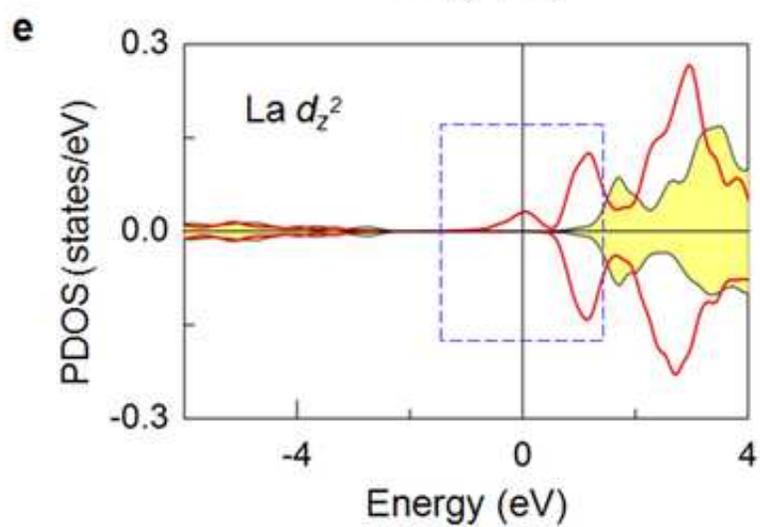
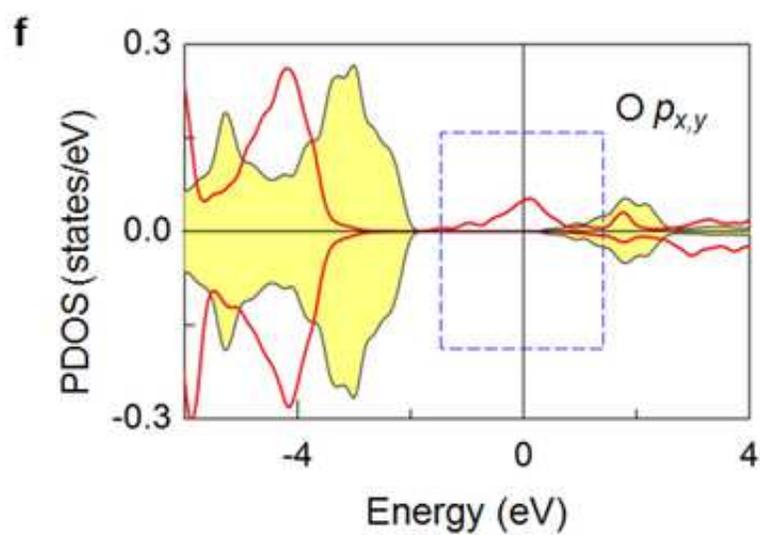
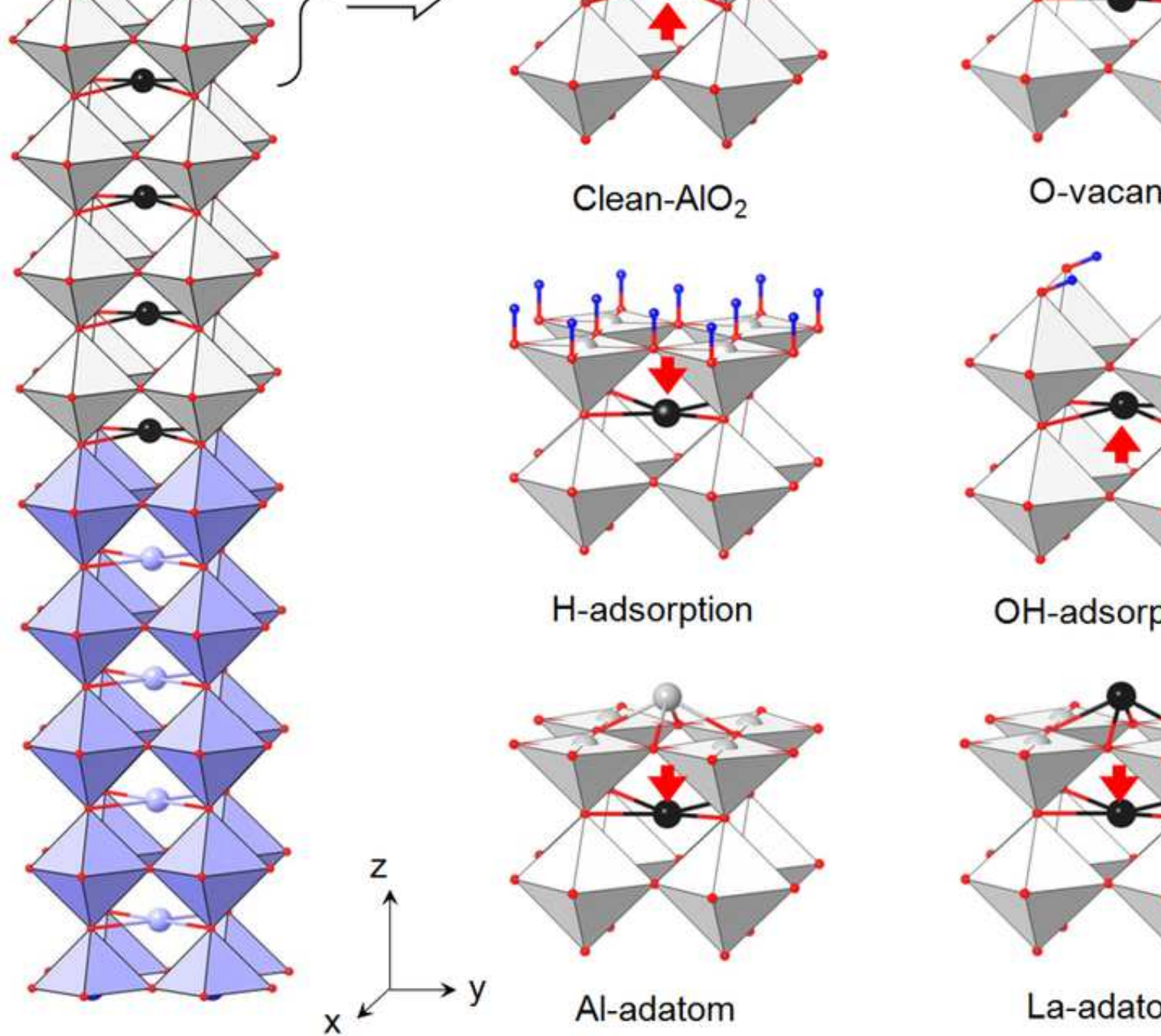
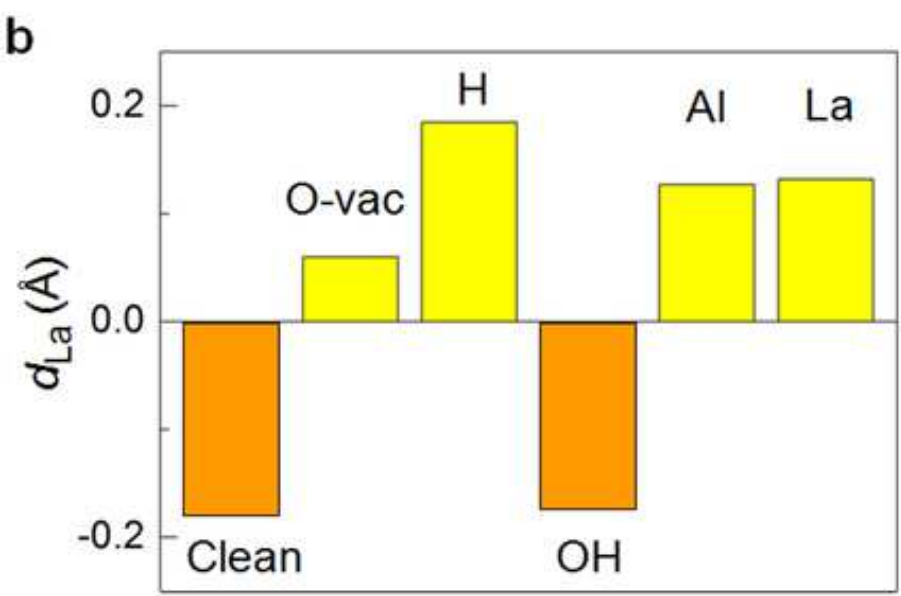
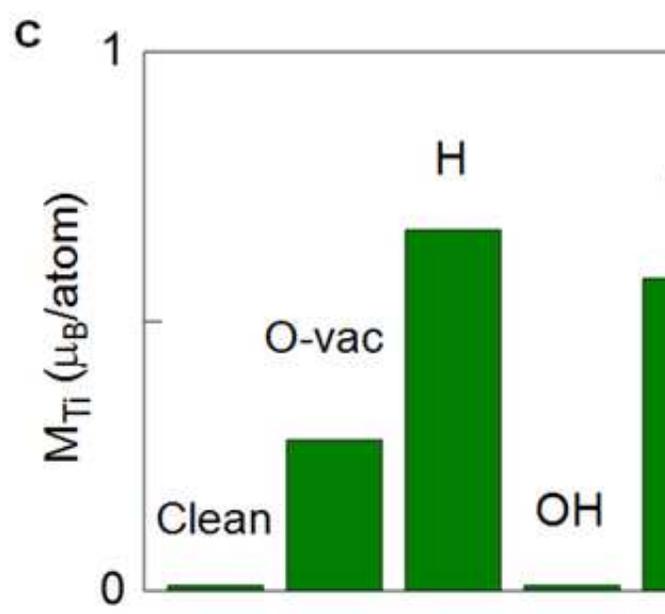
\end{document}